\documentclass[aps,prb,twocolumn,groupedaddress,amsmath,amssymb, showpacs]{revtex4-1}

\usepackage{graphicx,graphics,color,epsfig}
\usepackage{amsmath}
\usepackage{amssymb}
\usepackage{amsfonts}
\usepackage{bm}
\usepackage{times,xspace}
\usepackage{color}
\usepackage{array}
\usepackage{multirow}
\DeclareGraphicsExtensions{.pdf,.png,.jpg}

\begin{document}

\title{Intrinsic Large Gap Quantum Anomalous Hall Insulators in La$X$ ($X$=Br, Cl, I)}

\author{Kapildeb Dolui, Sujay Ray, Tanmoy Das}
\affiliation{Department of Physics, Indian Institute of Science, Bangalore, 560012, India.}
\date{\today}

\begin{abstract}
We report a theoretical prediction of a new class of bulk and intrinsic quantum Anomalous Hall (QAH) insulators La$X$ ($X$=Br, Cl, and I) via relativistic first-principle calculations. We find that these systems are innate long-ranged ferromagnets which, with the help of intrinsic spin-orbit coupling, become QAH insulators. A low-energy multiband tight binding model is developed to understand the origin of the QAH effect. Finally integer Chern number is obtained via Berry phase computation for each two-dimensional plane. These materials have the added benefit of a sizable band gap of as large as $\sim$ 25 meV, with the flexibility of enhancing it to above 75 meV via strain engineering. The synthesis of La$X$  materials will provide the impurity-free single crystals and thin-film QAH insulators for versatile experiments and functionalities.
\end{abstract}

\maketitle

\section{Introduction}
The discovery of quantum Hall effect in 1980 (Ref.~\cite{QHE}) has widely encouraged the study of topological invariants in solid state systems.\cite{TKNN,Haldane,Halperin,TIreviewCK,TIreviewSCZ,TIreviewTD} This has also contributed to the inception of many new physical concepts such as Thouless-Kohmoto-Nightingale-den Nijs (TKNN) invariants,\cite{TKNN} composite fermions,\cite{JainCF} and also raised the hope for the realization of many non-Abelian `particles' in the condensed matter setups.\cite{TIreviewCK,TIreviewSCZ} Subsequently, quantum spin-Hall (QSH)\cite{QSHE} and quantum anomalous Hall (QAH)\cite{QAHE,QAHEV} states were realized in which Hall current arises in the absence of external magnetic field. In both cases transverse spin current arises from the chiral spin-orbit entanglement with and without time-reversal symmetry, respectively. In these two subclasses, density-functional theory (DFT) has taken the preceding role in predicting suitable materials, which were followed by experimental realizations. 

The fascinating interplay between topological, magnetic and transport properties of QAH effects has tremendous scientific and technological interests. Therefore, the discovery of single-crystal QAH insulators with intrinsic long-range magnetization is considered imperative. Most of the search for QAH materials have so far been within the two-dimensional (2D) QSH, or thin-films of topological insulators. The idea here is to suppress one of their chiral spin current by inducing ferromagnetic (FM) state through impurity, doping or proximity effect. This guideline has led to the predictions of a variety of materials within the magnetically doped graphene,\cite{grapheneafm} QSH systems,\cite{HgCr} thin films of topological insulators,\cite{QAHTITh} quantum well states of Weyl semimetal HgCr$_2$Se$_4$,\cite{HgQW} and related materials (a comprehensive list of the predicted QAH materials is provided in Table~\ref{Tab:Mat}). Thin films of topological insulators are synthesized to date to exhibit QAH effect with magnetic impurities.\cite{QAHE,QAHEV} However, so far there is no single crystal with innate magnetic order and spin-orbit coupling (SOC) know to be QAH insulator.

We begin by considering the vast possibility of magnetic materials with strong SOC, and narrow down the list by eliminating the choices via crystal structure selection, absence of continuous magnetic transition, or impurity prone lattice etc. By using density-functional theory (DFT) calculations, we find that the FM La$X$ ($X$=Br, Cl, and I) is a layered material class which exhibits band crossing between different orbitals and opposite spins at a momenta contour near the Fermi level without SOC.\cite{WOS} As SOC is switched on, the `inverted' insulating band gap opens in LaBr, LaCl and LaI, but not in LaF, endowing the former systems into the uncharted territory of intrinsic QAH insulators. To understand and characterize these properties, we have also formulated a realistic tight-binding model and estimated the QAH index via Berry curvature computation. LaBr and LaCl are also found to be flexible for enhancing their band gap via small strain. Much like how Bi- and Sb- based binary topological insulators provided the breakthrough for the discovery and predictions of many related materials and their ternary and quaternary variants, our prediction provides a guiding principle for manipulating and expanding the `bulk' and intrinsic QAH insulator classes in related materials. 

The reminder of this article is organized as follows. In Sec.~\ref{Sec:Mat}, we describe the materials selection process for the QAH insulators. The corresponding first-principles band structure calculation for La$X$ family is performed in Sec.~\ref{Sec:DFT}. Then we develop a material specific two band tight-binding model for La$X$ with SOC and FM order in Sec.~\ref{Sec:TB}. The Berry phase and Chern number calculations are presented in Sec.~\ref{Sec:Berry}. We have also demonstrated the presence of edge states in Sec.~\ref{Sec:Edge}. Strain enhancement of the bulk gap is shown in Sec.~\ref{Sec:Strain}. Finally, we conclude in Sec.~\ref{Sec:Conclusion}. In Appendix~\ref{SecA:Band}, we present the details of the band progression from paramagnetic state to the FM state with and without SOC. The full derivation of the tight-binding model is provided in Appendix~\ref{SecA:TB}. A list of prior predicted and experimentally realized materials with the corresponding tuning parameter and band gap is created in Table~\ref{Tab:Mat}.

\section{Materials selection}\label{Sec:Mat}

Candidate materials for bulk QAH insulators may be sought in compounds where (1) magnetism and SOC can be intrinsically present; (2) the magnetic transition is continuous with temperature, or otherwise the magnetic interactions would often not be long-ranged; (3) a stable non-high symmetric lattice structure which would have higher propensity towards band inversion\cite{TIreviewTD}; and (4) single crystal growth would be feasible with clear surface state, where dopant or impurity scattering can be made sufficiently small to achieve large values of transport properties. 
   
Simple candidates may include $d$-electron ferromagnets in the heavy-element compounds, or intercalated with heavy-elements for sufficiently large SOC. But these systems often have cubic or tetragonal lattice, and in some cases, lack inversion symmetry which sometimes reduces the band inversion strength.\cite{TIreviewTD} Similar complication arises in simple $d$-electron antiferromagnets in which, moreover, the magnetic transitions are sometimes not continuous. In addition, much of these systems suffer from the problems of impurity and low transport properties.

More promising candidates can be expected within the actinide and lanthanide compounds. Actinides have larger SOC strength. But due to larger bandwidth of 5$f$ states, the interaction strength decreases and they often escapes magnetism. More importantly, radioactive materials suffer from radiation induced defects. We then consider lanthanide based rear-earth elements. Their low-energy spectrum remain poised between the $d$- and $f$- electronic configurations, which make them versatile for commencing diverse materials properties. Among the rear-earth elements, lanthanum (La) provides the common element for materials with properties as diverse as half-Huesler topological insulators,\cite{LaTI1,LaTI2} colossal magnetoresistance,\cite{LaCMR} metal-insulator transition,\cite{LaMIT} as well as high-temperature superconductivity.\cite{LaTc} Therefore, La is a good candidate element to search for large gap intrinsic QAH effect.   

\begin{figure}[t]
\centering
\rotatebox[origin=c]{0}{\includegraphics[width=0.99\columnwidth]{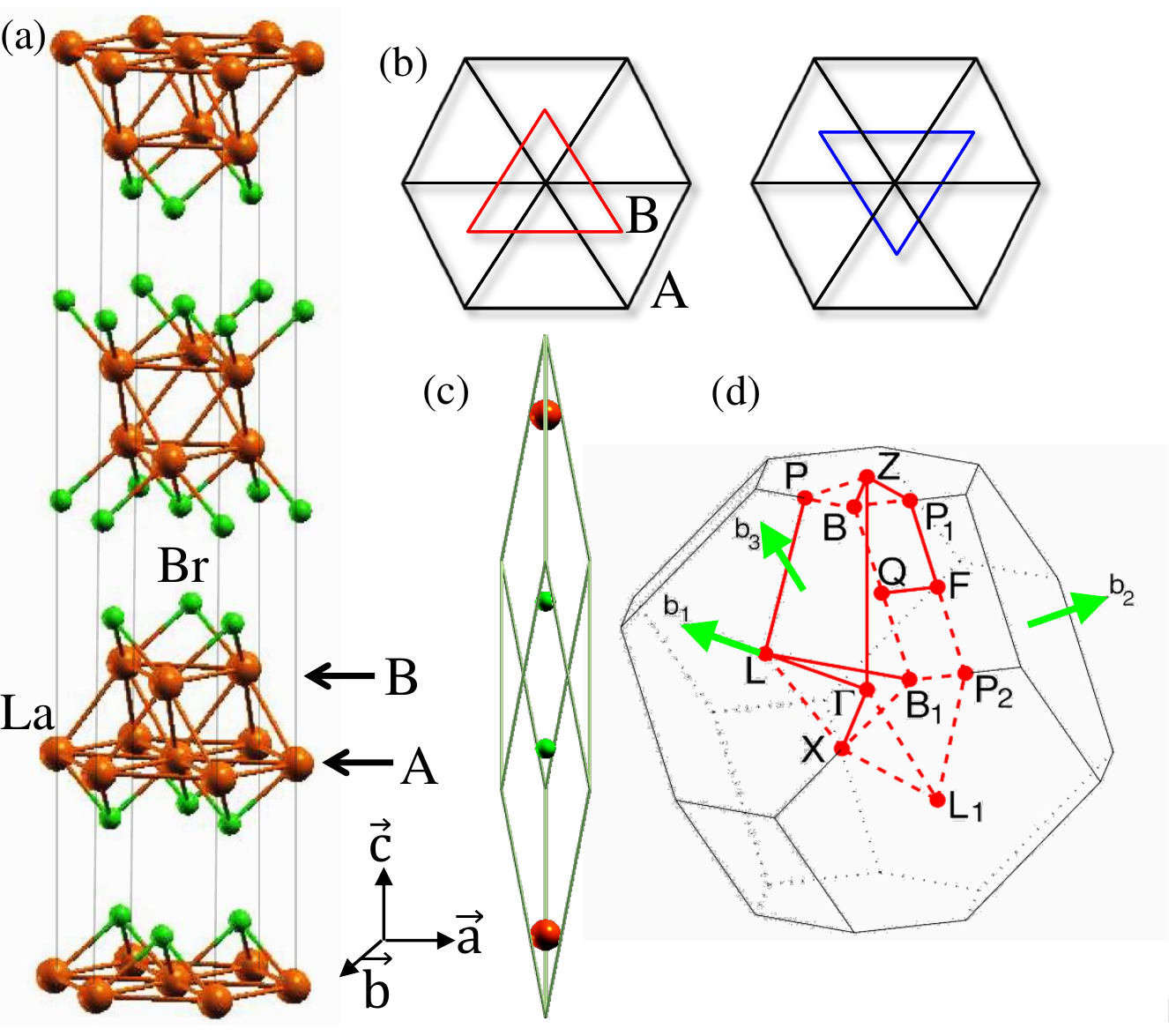}}
\caption{(Color online) (a) Crystal structure of LaBr, with two inequivalent La atoms forming a bilayer, defined by a hexagonal layer with a triangle layer on its top or bottom plane, respectively. (b) Top view of two bilayers. The middle bilayer in (a) is same as other bilayers, but shifted in the $a$-$b$ plane. (c) Primitive rhombohedral unit cell of LaBr. (d) Brillouin zone with space group of $R\bar{3}m$. 
}\label{fig1}
\end{figure}

For the anion element, we consider halogen elements because of their high electronegativity which enables strong chemical bonding with La cations. Furthermore, halogen elements form very weak bonds within the diatonic ($X_2$) element, raising the possibility for obtaining layered systems with topological consequences. Our search by DFT calculations indeed yielded in a stable compounds La$X$ ($X$=Br, Cl, and I), which belong to the rhombohedral crystal structure with the space group $R{\bar 3}m$ and point group $D^5_{3d}$. As shown in Fig.~\ref{fig1}(a), the structure is characterized by a sequence of triangular network layers composed of La and $X$ atoms. Each bilayer has two inequivalent La atoms (denoted by `A' and `B' layers) arranged in patterns as demonstrated in Fig.~\ref{fig1}(b). All such bilayers are related to each other by inversion symmetry along the $c$-axis, and the inter-bilayer interaction arises from the weak van-der Waals interaction and is thus considerably small. 

\section{Band structure calculations}
\subsection{First-principles calculation}\label{Sec:DFT}
Electronic structure calculations are carried out using DFT within the generalized gradient approximation (GGA)\cite{PBE} as implemented within the VASP package\cite{VASP}. Projected augmented-wave (PAW)~\cite{PAW} pseudopotentials are used to describe core electrons. GGA+U method is used to deal with the strong correlations in these materials with the standard value of $U$ = 8 eV on the correlated La-5$d$ orbitals.\cite{LaBr3}. We have also explored a large range of $U$=0 to 8~eV, and the results remain characteristically the same. Moreover, the band structure is reproduced with other functionals including the local density approximation (LDA) + U, the van der Waals (vdW)-density functional (DF) [optB86b-vdW functional] method\cite{vdW} and the hybrid functional HSE06~\cite{HSE}. The conjugate gradient method is used to obtain relaxed geometries. Both atomic positions and cell parameters are allowed to relax, until the forces on each atom are less than 1 meV/\AA. The optimized lattice vectors and atomic positions are listed in Ref.~\cite{SI}. The kinetic energy cutoff of 650 eV and a $k$-mesh grid of 8$\times$8$\times$8 are used in the self-consistent calculations. In order to analyze the stability of lattice dynamics, force constants are calculated for a 2$\times$2$\times$2 super-cell within the framework of density functional perturbation theory. Subsequently, the phonon dispersions are calculated using Phonopy package\cite{phonopy}. The formation enthalpy of LaX compound is calculated as: $H_{\rm f}^{\rm DFT}$ = E$_{\rm tot}$(LaX) - E$_{\rm tot}$(La) - E$_{\rm tot}(X)$, where E$_{\rm tot}$(LaX), E$_{\rm tot}$(La) and  E$_{\rm tot}(X)$ are the total energy per formula-unit of La$X$,  La and $X$ respectively, in their bulk form. Both the phonon dispersion and formation energy show that La$X$ compounds are stable. Finally, given the valency of La ion being three, the natural compound to expect is La$X_3$.\cite{LaBr3} Therefore, relative stabilities of the multiphase system are analyzed by comparing their formation enthalpies. Here, the formation enthalpy of La$X$ is found to differ only slightly from that of La$X_3$, and thus the former sample can also be grown easily. See Ref.~\cite{SI} for further details of the DFT calculation and the results with various functionals, different values of $U$, 2D band plots, and others.

Both with and without SOC, FM order is found to be the stable ground state for all the present systems, with magnetic moment of $m= 0.8 \mu_B$ per La atom. The FM order, oriented perpendicular to the La-layers, is mediated by direct exchange interactions. Total energy of FM ground states is lower by 0.22 and 0.21 eV/unit cell than the corresponding nonmagnetic and antiferromagnetic states, respectively.  

\begin{figure}[t]
\centering
\rotatebox[origin=c]{0}{\includegraphics[width=0.90\columnwidth]{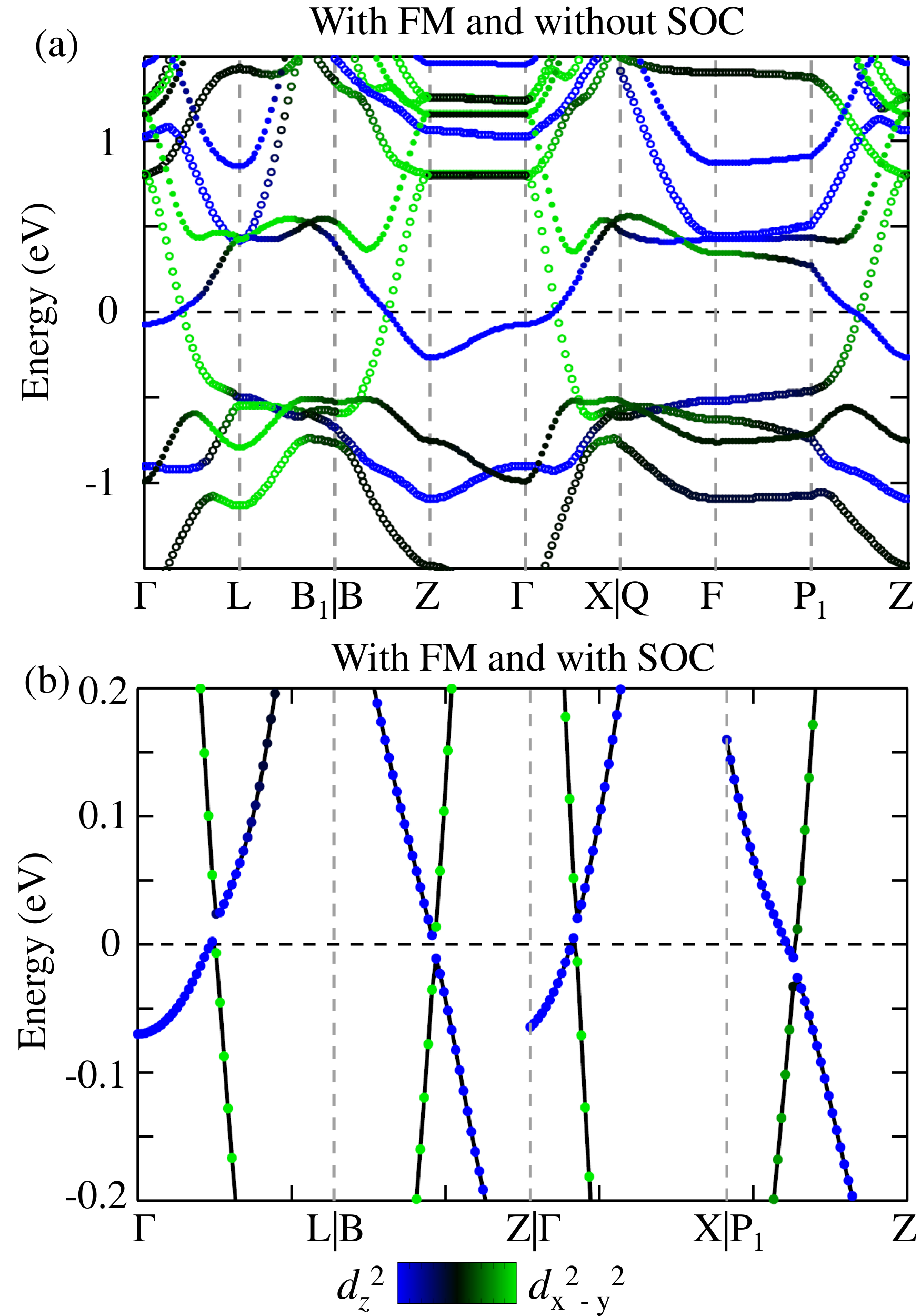}}
\caption{(Color online) (a) Band structure of LaBr in the FM state without SOC. The horizontal dashed line gives the Fermi level. Blue and green colors indicate $d_{z^2}$ and $d_{x^2-y^2}$ orbitals weight of the La atoms, while the black color is the weight of the other orbitals. Open and filled symbols represent spin up and spin down states, respectively. (b) Low energy view of the same band structure, but with SOC and plotted only along representative directions where the band gap opens across the Fermi level. The overall band structure at other directions and higher energy is similar to that of (a). Black solid lines are the guide to the eyes, where symbols are the calculated points.
}\label{fig2}
\end{figure}

The band progression of the representative compound LaBr is shown in Fig.~\ref{fig2} (also see Fig.~\ref{figA:Band}). La has the 4$f$ electrons fully filled, and its local moment is not often accessible to the conduction electrons. Magnetic moment arises in the half-filled 5$d$-shell as a result of large Hund's rule coupling. The local crystal field environment of the La ion leads to large splitting between the $e_g$ and $t_g$ states, and two partially filled $e_g$ orbitals osculate the Fermi level, with an overlapping semi-metallic like band structure.

In the FM state, strong Hund's coupling leads to spin splittings of the $d_{x^2-y^2}$ and $d_{z^2}$ orbitals. The two majority and minority spin states from two different orbitals cross each other across the Fermi level, creating an inverted band structure near $E_{\rm F}$, as shown in Fig.~\ref{fig2}(a) for several high-symmetric directions. Such contour of inverted band crossing is often called Dirac line nodes, or nodal Weyl ring \cite{WOS,Aji,Kane,Mullen} The average FM exchange splitting can be estimated as $\Delta E=2Jm=0.8$eV. By the value of the exchange coupling $J=0.5$~eV, we can estimate the FM Currie temperature to be as high as 390~K.

Since the Weyl ring occurs between different spin states from different orbitals, as the SOC is turned on, all $k$-points on the Weyl rings become gapped, and the system enters into an insulator state, see Fig.~\ref{fig2}(b) (also see Ref.~\cite{SI} for 2D band plots and density of states). We obtain a band gap of 14~meV for LaCl, and 22~meV for LaBr and 25~meV for LaI without tuning. Tiny semimetallic Fermi pockets are present for the calculations with relaxed lattice structure, which are lifted from $E_{\rm F}$ with small strain (see Fig.~\ref{fig4}). With strain and/or pressure, the insulating band gap can be enhanced to above 75 meV. As the band gap opens at the inverted bands crossing points via SOC, QAH effect can be expected in such systems.

\begin{figure}[t]
\centering
\rotatebox[origin=c]{0}{\includegraphics[width=0.80\columnwidth]{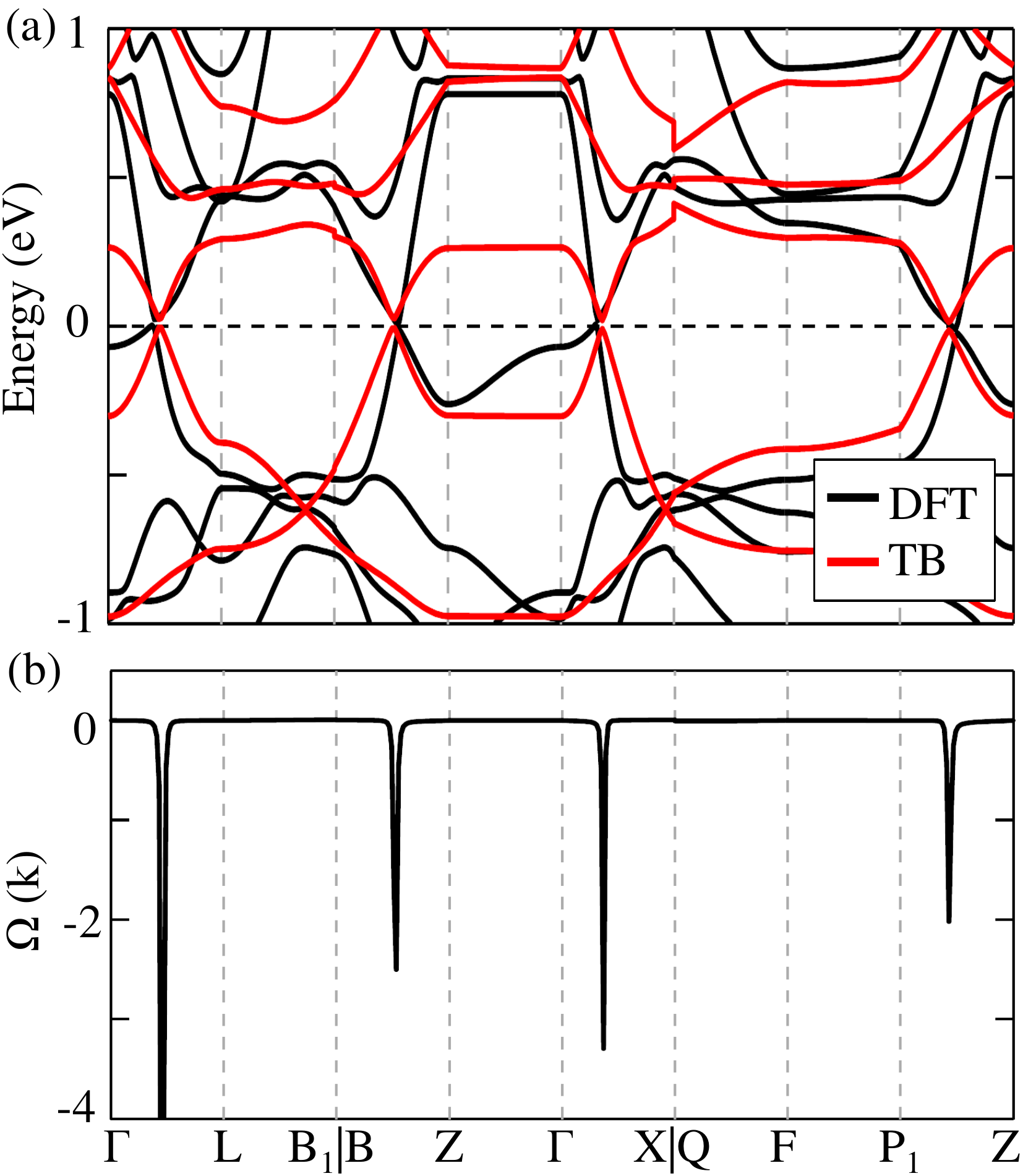}}
\caption{(Color online) (a) Tight binding band structure (red lines), overlayed on top of the DFT band structure (black lines) of LaBr with FM order and SOC. (b) Computed Berry curvature along the same high-symmetry directions as in (a), exhibiting spikes at the momenta where SOC gaps the FM Weyl points.   
}\label{fig3}
\end{figure}

\subsection{Tight-binding Hamiltonian and Berry phase computation}\label{Sec:TB}
Guided by the DFT band structure, we derive a realistic tight-binding model by including the two low-lying $e_g$ orbitals. The FM ordering is introduced by the mean-field Heisenberg term, in which we allow exchange couplings to be orbital dependent ($J_1$, $J_2$). The inter-orbital SOC is restricted to be nearest neighbor, which turns out to have a form of $\lambda_{k}=\lambda(\sin{k_x}+i\sin{k_y})$, where $\lambda$ is the SOC coupling strength. Two adjacent La layers are inequivalent (denoting by `A' and `B'), in which each La atom is surrounded by six other La atoms in a hexagonal plane, and three La atoms in a trigonal plane are placed at $\sim c/10$ distance along the $z$-axis (see Fig.~\ref{fig1}(a)). If we set our multiband spinor for each layer as $\psi_{k}^{A/B}=\left(|d_{x^2-y^2}^\uparrow\rangle,~|d_{z^2}^\uparrow\rangle,~|d_{x^2-y^2}^\downarrow\rangle,~|d_{z^2}^\downarrow\rangle\right)$, the bulk Hamiltonian can then be expressed as $H_{A}\left( k\right)$ =
\begin{eqnarray}
\left(
 \begin{array}{ c c c c }
 \mathcal{\xi}_{1k}+J_{1}m & \mathcal{\xi}_{3k}        &  0                      & \lambda_{k}\\
 \mathcal{\xi}_{3k}^{*}   & \mathcal{\xi}_{2k}-J_{2}m  & \lambda_{k}                      & 0\\
 0                      &                   \lambda_{k}^* & \mathcal{\xi}_{1k}-J_{1}m & \mathcal{\xi}_{3k}\\
 \lambda_{k}^*                  &                       0 & \mathcal{\xi}_{3k}^{*}   & \mathcal{\xi}_{2k}+J_{2}m
 \end{array} \right).
\label{Ham}
\end{eqnarray}
For the lattice structure depicted in Fig.~\ref{fig1}, we obtain the non-interacting dispersions $\xi_{jk}$ for both the intra- ($j$=1, 2) and inter-orbitals ($j$=3) as  $\xi_{jk}=2t_j^{1}\cos k_{x}a+4t_j^{2}\cos \frac{k_{x}a}{2}\cos \frac{\sqrt{3}k_{y}a}{2}-\mu_j  $. Here $t_j^i$ are the nearest neighbor tight-binding hopping parameters for $i^{\rm th}$ orbital. $\mu_1=\mu$ is the chemical potential, $\mu_2=\mu+\Delta$, includes both the chemical potential and the onsite energy difference between two orbitals, and $\mu_3=0$. $a$ and $c$ are the in-plane and out-of plane lattice constants. Due to symmetry consideration, the Hamiltonian for the `B' plane is $H_B=H_A$. Finally, nearest neighbor hopping terms between `A' and `B' layers turn out to be 
\begin{eqnarray}
\xi_{jk}^{AB} = t_j^3e^{-i\left(\frac{k_{y}a}{\sqrt{3}} + \frac{k_{z}c}{10}\right)  } + 2t_j^{4}\cos{\frac{k_{x}a}{2}} e^{i\left(\frac{k_{y}a}{2\sqrt{3}} - \frac{k_{z}c}{10}\right)  }.
\end{eqnarray}
Here also, $j$ stands for the same intra- ($j=$1,2) and inter-orbital ($j$=3) hoppings. The parameter values are deduced by fitting to the DFT dispersion as shown in Fig.~\ref{fig3}(a). Our fitting yields $J_1$ = -0.43~eV, $J_2$ = -0.69~eV, $\lambda$ = 20 meV, which are close to the DFT values listed above. The other parameters are given in Appendix~\ref{SecA:TB}. The overall fitting is reasonably good and the TB band structure captures the important topological properties including band inversion, nodal Weyl ring, and the insulating gap formation with SOC in accordance with the DFT calculation. 

\section{Topological properties of La$X$}
\subsection{Berry curvature and Chern number}\label{Sec:Berry}
Near the Weyl ring, the same Hamiltonian can be reduced to that of the topological insulator Hamiltonian.\cite{TIreviewSCZ} Since band inversion has already occurred between opposite spin states, SOC induced insulating gap renders a QAH insulators.\cite{grapheneafm,QAHTITh}. Due to the quasi-2D nature of the La-bilayes, the topological invariant for the QAH can be estimated from the Berry connection.\cite{grapheneafm,QAHTITh,BerryQAH} We have computed Berry curvature, $\Omega (k_x,k_y)$ at constant $k_z$, by using the Kubo formula:
\begin{eqnarray}
\Omega (k)&=&\hbar ^{2} \sum_{n\neq n'}\frac{{\rm Im}[\langle nk|v_{x}|n'k\rangle \langle n'k|v_{y}|nk\rangle]}{(E_{nk}-E_{n'k})^{2}}\nonumber\\
&&\times\left[ n_{f}(E_{nk})- n_{f}(E_{n'k})\right].
\end{eqnarray}
Here $v_{x}$ and $v_{y}$ are velocity operators in $x$, and $y$ directions, respectively, defined as $v_{x}=\frac{1}{\hbar}\frac{\partial H}{\partial k_{x}}$ and $v_{y}=\frac{1}{\hbar}\frac{\partial H}{\partial k_{y}}$. $\frac{\partial H}{\partial k_{x/y}}$ represents derivative of each element of the Hamiltonian matrix. $|nk\rangle$ and $|n'k\rangle$ are eigenvectors of H with eigenvalue $E_{nk}$ and $E_{n'k}$, respectively. And  $n_{f}(E_{nk})$ is the Fermi occupation number at energy $E_{nk}$. The obtained Berry curvature is plotted along the representative high symmetry directions (for different directions, the corresponding $k_z$ values are kept constant), in Fig.~\ref{fig3}(c) along the high-symmetry directions. $\Omega ({\bf k})$ exhibits strong spike-like anomaly at the gapped Weyl points, signaling the presence of Hall plateau.\cite{grapheneafm}  

Finally, we calculate the Hall conductance $\sigma_{xy}$ which is related to the Chern number \textit{C} by $\sigma_{xy}=C\dfrac{e}{h^{2}}$ where the Chern number \textit{C} is given by 
\begin{equation}
 C = \dfrac{1}{2\pi}\int_{BZ} \Omega (k)d^{2}k
 \end{equation}
Each 2D La-layer constitutes a hexagon. The Chern number is calculated by summing the Berry curvature on each $k_z$ planes which comes out to be $-1$, signaling the presence of QAH effect. The result is valid even in the semimetallic state as long as the Fermi surface pocket is small enough, and a single valued Chern number can be defined for the filled bands. 

\subsection{Edge state}\label{Sec:Edge}
\begin{figure}[h!]
\centering
\rotatebox[origin=c]{0}{\includegraphics[width=0.8\columnwidth]{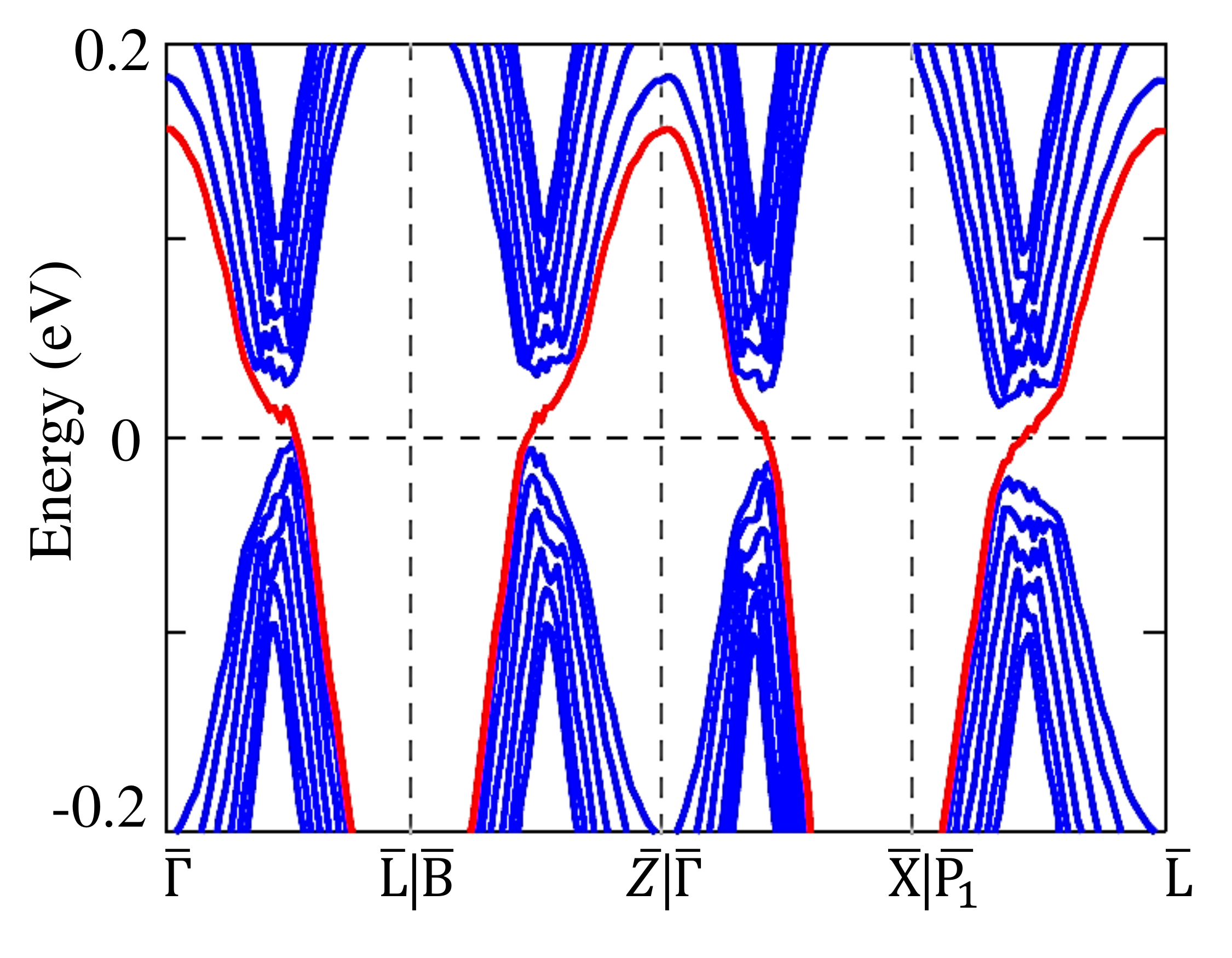}}
\caption{Edge states (red lines) along representative high-symmetry directions where SOC opens a gap in Fig.~\ref{fig2}(b).}
\label{fig4}
\end{figure}
We also calculate the edge state for various representative cuts and the results are shown in Fig.~\ref{fig4}. For all cases, we have used 8 La-bilayers, and the cut is chosen in such a way that the direction of plotting is fully periodic while the perpendicular direction has open boundary condition. For example, along the $\bar{\Gamma}\rightarrow {\bar {\rm L}}$ direction, we have made the supercell along $b$-direction, with $k_x$ and $k_z$ remaining the good quantum number. Similar consideration is taken for all other directions. We find that the magnetically polarized edge state adiabatically connects both the bulk valence and conducting bands, following the topological criterion of the QAH state.\cite{TIreviewCK,TIreviewSCZ,TIreviewTD}

\section{Strain engineering and band gap enhancement}\label{Sec:Strain}

\begin{figure}[t]
\centering
\rotatebox[origin=c]{0}{\includegraphics[width=0.75\columnwidth]{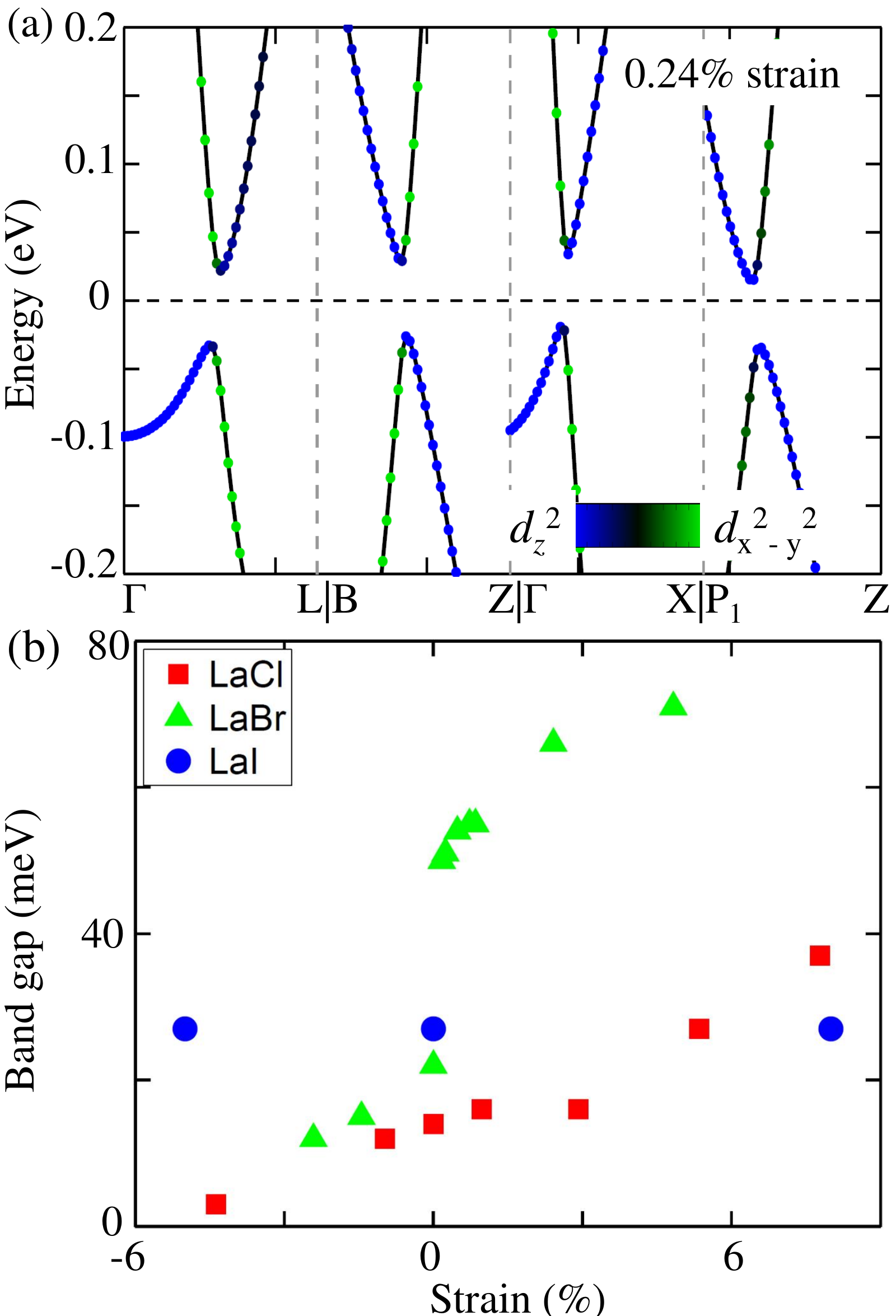}}
\caption{(Color online) Strain engineering and band gap tunning. (a) Insulating band structure of LaBr in the FM + SOC state with only 0.24\% uniaxial strain. (b) Strain induced band gap evolution for all three QAH systems.
}\label{fig5}
\end{figure}
Strain is a straightforward and efficient method to tune band gap. We find that uniaxal strain can increase the band gap of LaBr and LaCl systems up to two to three times its value in the relaxed coordinates. Moreover, only a very small amount of strain is necessary to remove the states from $E_{\rm F}$ at all momenta as shown in Fig.~\ref{fig5}(a) for 0.24\% strain in LaBr. The full strain dependence of the band gap is plotted in Fig.~\ref{fig5}(b). It is interesting to see that the band gap of LaBr increases rapidly for very small positive strain, without undergoing any structural transition. For LaCl, the band gap tunning is very much monotonic with strain. On the other hand, LaI does not exhibit any considerable band gap change for this range of strain, presumably due to large ionic size of the I atom the effect of strain is reduced here. A comprehensive list of various QAH systems, their tuning mechanism, and the corresponding gap values is provided in Table~\ref{Tab:Mat}. In comparison to the existing materials which all require some sort of tuning to obtain QAH state, the present systems are not only intrinsic QAH insulators, but also provide an opportunity to obtain band gap larger than most of the earlier ones.
 
\section{Conclusion}\label{Sec:Conclusion}
Our prediction will help realize the long-sought goal where the QAH effect can be realized and utilized in single crystals. Moreover, the predicted band gap is large enough for many practical and experimental purposes, and larger than most of the earlier predicted systems. The layered structure of La$X$ compounds are easy to cleave to obtain topological surface states. These advantages will allow full access to the thermal and electric transport regime of the QAH edge states without having the problem of magnetic impurity or thin film growth. Furthermore, this work will accelerate the prediction and discovery of more intrinsic QAH insulators within the lanthanides and beyond.

Diverse other possibilities can be explored in intrinsic QAH insulators with long-ranged FM ordering. Goldstone mode, a collective zero-energy magnetic excitation, is a natural consequence of the continuous symmetry breaking, which can be expected to evolve with topological consequences in the QAH insulator.\cite{Goldstone} Chiral spin-current in the edge state can lead to skyrmions excitations.\cite{skyrmions} Owing to the heavy-electron mass of La electrons, other rear-earth substitution in the La site can bring in the Kondo physics, mixed valence state within the topological matrix. The surface states of the QAH insulator is also a potential host of axion particle.\cite{Axion} 
\\\\
{\bf Acknowledgments}
The work is facilitated by the Bardeen HPC cluster facility in the Physics Department of the Indian Institute of Science.

\appendix
\section{Full band progression with and without SOC or FM order}\label{SecA:Band}
\begin{figure}[t]
\centering
\rotatebox[origin=c]{0}{\includegraphics[width=1.0\columnwidth]{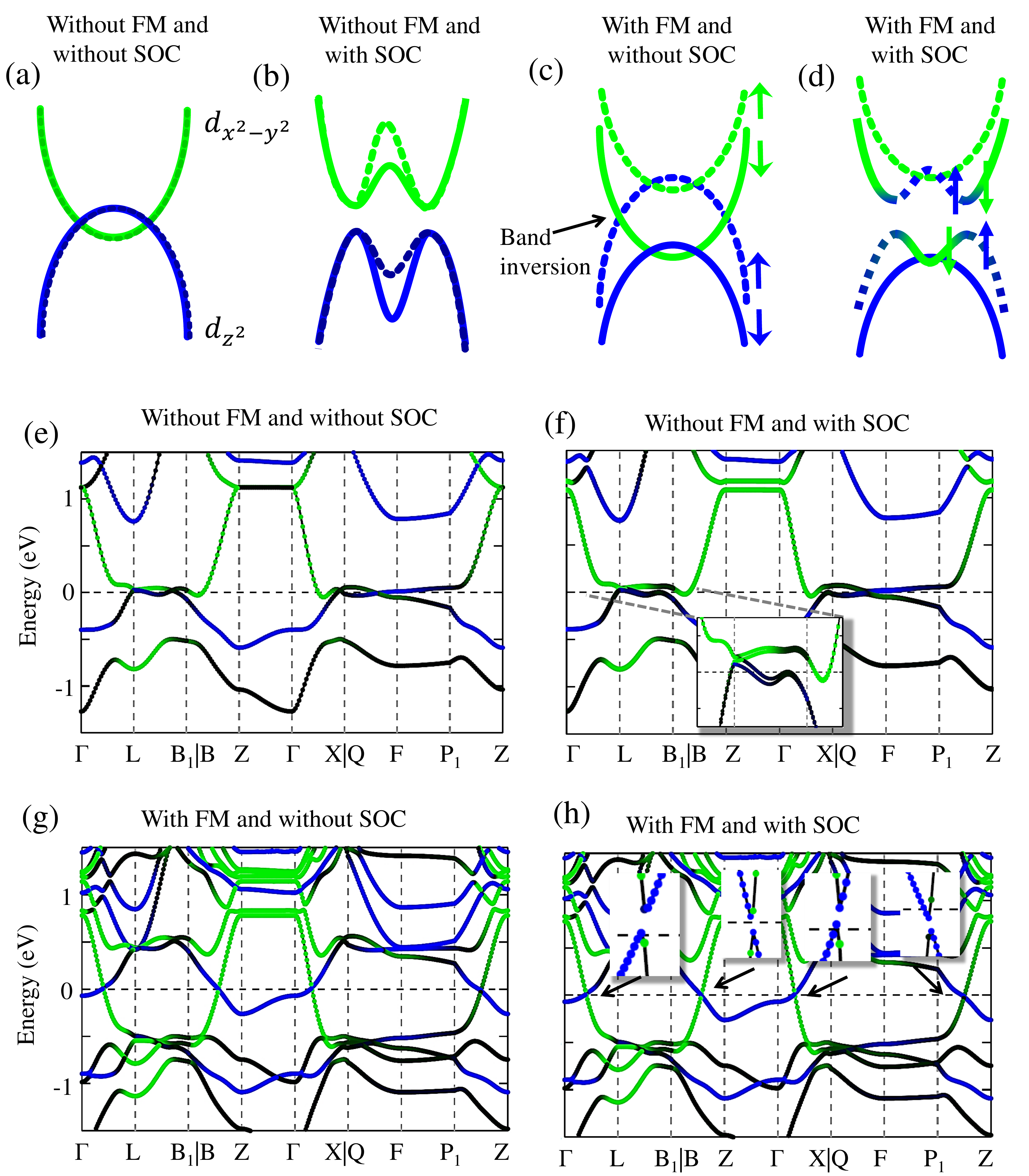}}
\caption{Band progression of La$X$ materials. (a-d) The upper panel schematically illustrates the band structure around the band crossings between two different orbitals near the Fermi level for four different cases. (a) Band structure in the absence of FM order and SOC. (c) Same as (a) but with SOC. This particular example is representative of the La$X$ materials where SOC splitting can be anisotropic across the band crossing points. (c) Bands with FM order and without SOC. (c) Once SOC is included, insulating gap form between different orbitals and spins. (e)-(h), DFT band structure for the four cases discussed in the upper panels in the same order. Various insets highlight the low-energy regions where SOC induces band gap.}
\label{figA:Band}
\end{figure}
In Fig.~\ref{figA:Band}, we give the details of the band progression of LaBr with and without SOC and FM. The band crossing and band inversion here occur at non-high-symmetric $k$-points, and thus undergoes band splitting due to SOC even in the presence of time-reversal and inversion symmetries. This is demonstrated from the DFT calculation as well in Fig.~\ref{figA:Band}(f). In the FM state, the majority and minority bands from different orbitals cross each others at the Fermi level which in fact sets the stage for the band inversion and insulating gap when SOC is turned on. The band crossings occurs at a contour of $k$-points, which all become gapped with SOC and a fully insulating gap indeed opens in these systems. The 2D band plots at various $k_z$-cuts as well as density of states are shown in Ref.~\cite{SI}.

\section{Further details of the Tight-binding model}\label{SecA:TB}
Here we expand the description of the tight binding (TB) model. First we describe the non-magnetic state. Then the effect of magnetism is considered within a mean-field treatment, and finally the spin orbit coupling (SOC) is included. The TB parameters are obtained by fitting the band structure to the corresponding DFT calculation. The structure of LaBr is given by two layers of 2D hexagonal lattices stacked along the $c$ direction, as shown in Fig.~\ref{fig1}. The in-plane TB Hamiltonian therefore remains the same for each layer, and we have only considered the nearest neighbor inter-layer hopping:
\begin{eqnarray}
H = \left[
\begin{array}{ c c }
H_{A}  & H_{AB}\\
H_{AB}^* & H_{B}
\end{array} \right].
\end{eqnarray}
If we choose our origin of reference at the center of hexagon (point `A') in Fig.~\ref{fig1}(b), then the coordinates of nearest neighbor lattice points are given by: 
\begin{eqnarray}
&&{\rm A}\equiv (0,0,0), \qquad {\rm B}\equiv (a,0,0), \qquad {\rm C}\equiv (\frac{a}{2},\frac{\sqrt{3}a}{2},0), \nonumber\\
&&{\rm D}\equiv (-\frac{a}{2},\frac{\sqrt{3}a}{2},0), \qquad {\rm E}\equiv (-a,0,0),\nonumber\\
&&{\rm F}\equiv (-\frac{a}{2},-\frac{\sqrt{3}a}{2},0), \qquad {\rm G}\equiv (\frac{a}{2},-\frac{\sqrt{3}a}{2},0), \nonumber\\
&&{\rm H}\equiv (0,-\frac{\sqrt{3}a}{3},-\dfrac{c}{10}), \qquad {\rm I}\equiv (\frac{a}{2},\frac{\sqrt{3}a}{6},-\dfrac{c}{10}),\nonumber\\
&&{\rm J}\equiv (-\frac{a}{2},\frac{\sqrt{3}a}{6},-\dfrac{c}{10}).\nonumber
\end{eqnarray}
The effective two orbital Hamiltonian for each layer (say, A layer) is given by,
\begin{eqnarray}
H_{A}&=&\sum_{k,i=1,2,\sigma}\xi_{ik}c^{\dagger}_{ik\sigma}c_{ik\sigma}+\sum_{k,\sigma}\xi_{3k}c^{\dagger}_{1k\sigma}c_{2k\sigma}\nonumber\\
%
&&+\sum_{k,i,\alpha,\beta}J_{i}mc^{\dagger}_{ik\alpha}{\bf{\sigma}}^z_{\alpha \beta}c_{ik\beta}
+\sum_{k,i,j\neq i,\alpha,\beta\neq \alpha}\lambda_{k} c^{\dagger}_{ik\alpha}c_{jk\beta}.\nonumber\\
\end{eqnarray}
$c_{jk\sigma}$ is the destruction operator for the $j^{\rm th}$ orbital ($j=1$ for $d_{x^2-y^2}$ and 2 for $d_{z^2}$ orbital), $k$ is the crystal momentum index and $\sigma$ is the spin index which can take two values as $\uparrow$ for up spin and $\downarrow$ for down spin. The first term of the Hamiltonian represents intra orbital hopping; the second term is the interorbital hopping; the third term is the Heisenberg term; and the last one is the SOC term. $\xi_{ik}$ are intraorbital band dispersions for $d_{x^2-y^2}$ and $d_{z^2}$ orbitals, respectively and $\xi_{3k}$ is the interorbital band dispersion. $J_{i}$ are Heisenberg couplings, and $m$ is spontaneous magnetization. $\sigma^z$ is the third Pauli matrix. $\lambda_k$ is nearest neighbor SOC. 

In the TB model only nearest neighbor hoppings are considered. For a single layer each lattice site has six nearest neighbors at the vertices of the hexagon surrounding the lattice site. Also due to the similar spatial symmetry of the two $e_g$ orbitals considered here, phase differences between different sites for the two orbitals come out to be the same, expect TB hopping parameters. For simplicity, we define all TB parameters as $t_j^i$, where subscript $j$ denotes different orbitals ($j=$ 1 and 2 for two intra-orbital terms, and 3 for the inter-orbital term), while the superscript $i$ differentiates them for different sites. Hopping from site `A' to site `B' yields $t_j^1e^{i{\bf k}.({\bf r}_{\rm B}-{\bf r}_{\rm A})}=t_j^1e^{ik_xa}$ where ${\bf r}_{\rm A}$ and ${\bf r}_{\rm B}$ are the positions of `A'  and `B' sites, and `a' is the in-plane lattice constant. All TB parameters for hoppings to `C', `D', `F', `G' sites from site `A' are equal to each other as the distances are the same, and we denote it by $t_j^2$. Therefore, the in-plane dispersion for both intra- and inter-orbitals becomes
\begin{eqnarray}
\xi_{jk}&=&t_j^{1}\Big[e^{ik_{x}a}+e^{-ik_{x}a}\Big]\nonumber\\
&&+t_j^{2}\Big[\big\{e^{i(k_{x}\frac{a}{2}+k_{y}\frac{\sqrt{3}a}{2})}+e^{-i(k_{x}\frac{a}{2}+k_{y}\frac{\sqrt{3}a}{2})}\big\}\nonumber\\
&&+\big\{e^{i(k_{x}\frac{a}{2}-k_{y}\frac{\sqrt{3}a}{2})}+e^{-i(k_{x}\frac{a}{2}-k_{y}\frac{\sqrt{3}a}{2})}\big\}\Big]-\mu_{j},\nonumber\\
&=&2t_j^{1}\cos k_{x}a+4t_j^{2}\cos \frac{k_{x}a}{2}\cos \frac{\sqrt{3}k_{y}a}{2}-\mu_{j}.
\end{eqnarray}
The last term $\mu_{j}$ encodes various contributions: Lets say $\mu_1=\mu$ is the chemical potential of the system, then $\mu_2=\mu+\Delta$, where $\Delta$ is the onsite energy difference of the second orbital from the first one. Since the inter-orbital hopping involves neither an onsite term, nor the chemical potential, so $\mu_3=0$. For intra orbital hopping of $d_{z^2}$ orbital, due to its azimuthal symmetry, both TB parameters are the same, i.e., $t_2^1=t_2^2$.

Now we consider Hamiltonian for layer `B'. It is easily recognized from the structure that layer B and layer A are the same. We set $H_{B}(k) = H_{A}(-k)$. Next we derive the Hamiltonian for inter-layer dispersion $\xi^{AB}_{jk}$. The distance between the `A' and `B' layers is about $c/10$. The nearest neighbor site to the `A' atom are `H', `I' and `J'. By doing a similar algebra as for the planer dispersion, we obtain the bare dispersions for the inter-layer terms as
\begin{eqnarray}
\xi_{jk}^{AB} = t_j^3 e^{-i\left(\frac{k_{y}a}{\sqrt{3}} + \frac{k_{z}c}{10}\right)  } + 2t_j^4\cos\left( \frac{k_{x}a}{2}\right) e^{i\left(\frac{k_{y}a}{2\sqrt{3}} - \frac{k_{z}c}{10}\right)}.\nonumber\\
\end{eqnarray}
For the same azimuthal symmetry of the the $d_{z^2}$ orbital, $t_2^3=t_2^4$.

If we choose the basis of our wave function for each layer $\psi_{k}^A$ as $\psi_{k}^A=\left( c_{1k\uparrow}\qquad c_{2k\uparrow}\qquad c_{1k\downarrow}\qquad c_{2k\downarrow}\right)$ then the corresponding Hamiltonian can be expressed in a $4\times 4$ matrix (and $8\times 8$ matrix for two layers) as $H_{A}\left( k\right) = \langle\psi^A_{k}|H|\psi^A_{k}\rangle$. We include SOC in the present lattice as evaluated by Haldane\cite{Haldane} to be $\lambda_k=\lambda\left( \sin k_{x}+i\sin k_{y}\right)$. Therefore, by including the FM order and the SOC, we get the final Hamiltonian for the intra-layer (`A') as $ H_{A}\left( k\right) =$
\begin{eqnarray}
 \left[
 \begin{array}{ c c c c }
 \xi_{1k}+J_{1}m & \xi_{3k}        & 0                      & \lambda_k\\
 \xi_{3k}^{*}   & \xi_{2k}-J_{2}m  &\lambda_k                      & 0\\
 0                      &                   \lambda_k^* & \xi_{1k}-J_{1}m & \xi_{3k}\\
 \lambda_k^*                  &                       0 & \xi_{3k}^{*}   & \xi_{2k}+J_{2}m
 \end{array} \right],\nonumber\\
\end{eqnarray}
and for the inter-layer as
\begin{eqnarray}
 H_{AB}\left( k\right) = \left[
 \begin{array}{ c c c c }
 \xi_{1k}^{AB}       & \xi_{3k}^{AB} & 0                     & 0\\
 \xi_{3k}^{AB*}  & \xi_{2k}^{AB}  & 0                     & 0\\
 0                     &                0 & \xi_{1k}^{AB}       & \xi_{3k}^{AB}\\
 0                     &                0 & \xi_{3k}^{AB*}  & \xi_{2k}^{AB}
 \end{array} \right].
 \end{eqnarray}
 
The eigenvalues of the 8$\times$8 Hamiltonian give energy dispersions. We have fitted the energy bands obtained from our model to DFT bands and the values of the parameters are (in eV): $t_1^1=-0.0553,~t_1^2=-0.0597,~\mu_{1}=-1.2345,~t_2^1=t_2^2=0.037,~\mu_{2}=-0.0662,~t_3^1=0.0522,~t_3^2=0.0499,~t_1^3=0.4471,~t_1^4=-0.0776,~t_2^3=t_2^4=0.177, ~t_3^3=-0.1904,~t_3^4=-0.045,~m=0.8,~J_{1}=-0.4251,~J_{2}=-0.6927$, and $\lambda$=0.02.

The TB bands fit well to the DFT bands in the regions where the corresponding weights of the $d_{x^2-y^2}$ and $d_{z^2}$ orbitals are dominant. When we include the effect of magnetization but no SOC (i.e., $J_i$ are finite, $\lambda=0$) each energy band splits into two magnetic bands and there are crossing of bands at a contour of momentum, reproducing the nodal Weyl ring. This signals the presence of a band inversion in the FM states. Once the SOC is turned on (i.e., both $J_i$ and $\lambda$ are finite), inverted band gap opens at the band crossing points.

 \begin{table*}[h]
 \begin{tabular}{|m{3cm}|m{4cm}|m{2cm}|m{2cm}|}
 \hline
 Material &  Type of tuning & Band gap (meV) & Reference (s)\\
 \hline
 
 \multirow{3}{*}{Graphene} & Sc, Mn, Fe, Cu adatoms & 2.5, 4.5, 5.5, 7.0 & \cite{grapheneSc1,grapheneSc2}  \\
 \cline{2-4}
 & BiFeO${}_{3}$ AFM insulator (proximity effect) & 1.1 (4 with strain) & \cite{grapheneafm} \\
 \cline{2-4}
 & $5d$ transition metals (Hf, Ta, W, Re, Os, Ir, Pt) doping and applied electric field & 20-80 & \cite{graphene5d} \\
 \cline{2-4}
 \hline
 \multirow{3}{*}{Silicene} & 3$d$ transition metal absorbants & $\le$30 & \cite{silV} \\
 \cline{2-4}
 & Co doping & $<1$ & \cite{silCo}\\
 \cline{2-4}
 & Applied electric field & $\sim$1$e^{-3}$ & \cite{silefield}\\
 \hline
 
 Bi$_2$Te$_3$, Bi$_2$Se$_3$, Sb$_2$Te$_3$ (thin film, QW) & Cr, Fe doping & $\le$ 50 & \cite{QAHTITh,crdopped,feo}\\
 \hline
 
FM insulators (MnSe, MnTe, EuS) & Heavy metals (Bi, Pb, Hg etc.) substitution & 1-142 & \cite{heavymet} \\
 \hline

FM insulators in nGdN/ nEuO/ SrO heterostructure & Heterostructure engineering & 3-30 & \cite{magsalt} \\
 \hline
 
2D Organic topological insulators & Mn doping &  9.5 & \cite{inorg} \\
 \hline
 
 InSb-based ($p$-$n$ junction QW) & Mn doping & 5 & \cite{InMn} \\
 \hline
 
HgTe QW & Mn doping & $\le$1$e^{-3}$ & \cite{HgQW} \\ 
 \hline
 
 $\mathrm{Hg}{\mathrm{Cr}}_{2}{\mathrm{Se}}_{4}$ & Thin film and QW &  $\le$1$e^{-3}$ & \cite{HgCr} \\
 \hline
 
Double perovskites (${\mathrm{Sr}}_{2}{\mathrm{FeMoO}}_{6}$, Ba$_2$FeReO$_4$, Sr$_2$CrWO$_6$, La$_2$MnIrO$_4$/LaAlO$_3$) & Heterostructure & 0.1-26 & \cite{Srbilayer,Labilayer} \\ 
 \hline
 
Oxides (LaNiO$_3$, SrTiO$_2$/SrIrO$_3$, Na$_2$Ir$_2$O$_7$) & Heterostructure & - & \cite{oxide1,oxide2,oxide3} \\
 \hline
 
Topological crystalline insulator & Thin-film and FM doping & - & \cite{TCI} \\
 \hline

La$X$ ($X$ = Cl, Br, I) & Intrinsic & 10-30 (40-75 with strain) & Present work \\
 \hline
\end{tabular}
\caption{List of predicted QAH systems with types of tunning and maximum band gap estimation. For the second last two systems, the band gap depends on the tight-binding parameter values, and the realistic values are not known. The chosen family of materials are representative, and there can be  more materials in these families, but the band gap is lesser.}\label{Tab:Mat}
 \end{table*}


\noindent
\clearpage
\newpage

\begin{widetext}

{\large{\bf Supplementary materials for ``Intrinsic Large Gap Quantum Anomalous Hall Insulators in La$X$ ($X$=Br, Cl, I)"}}

\begin{figure}[h!]
\centering
\fbox{\includegraphics[scale=0.5]{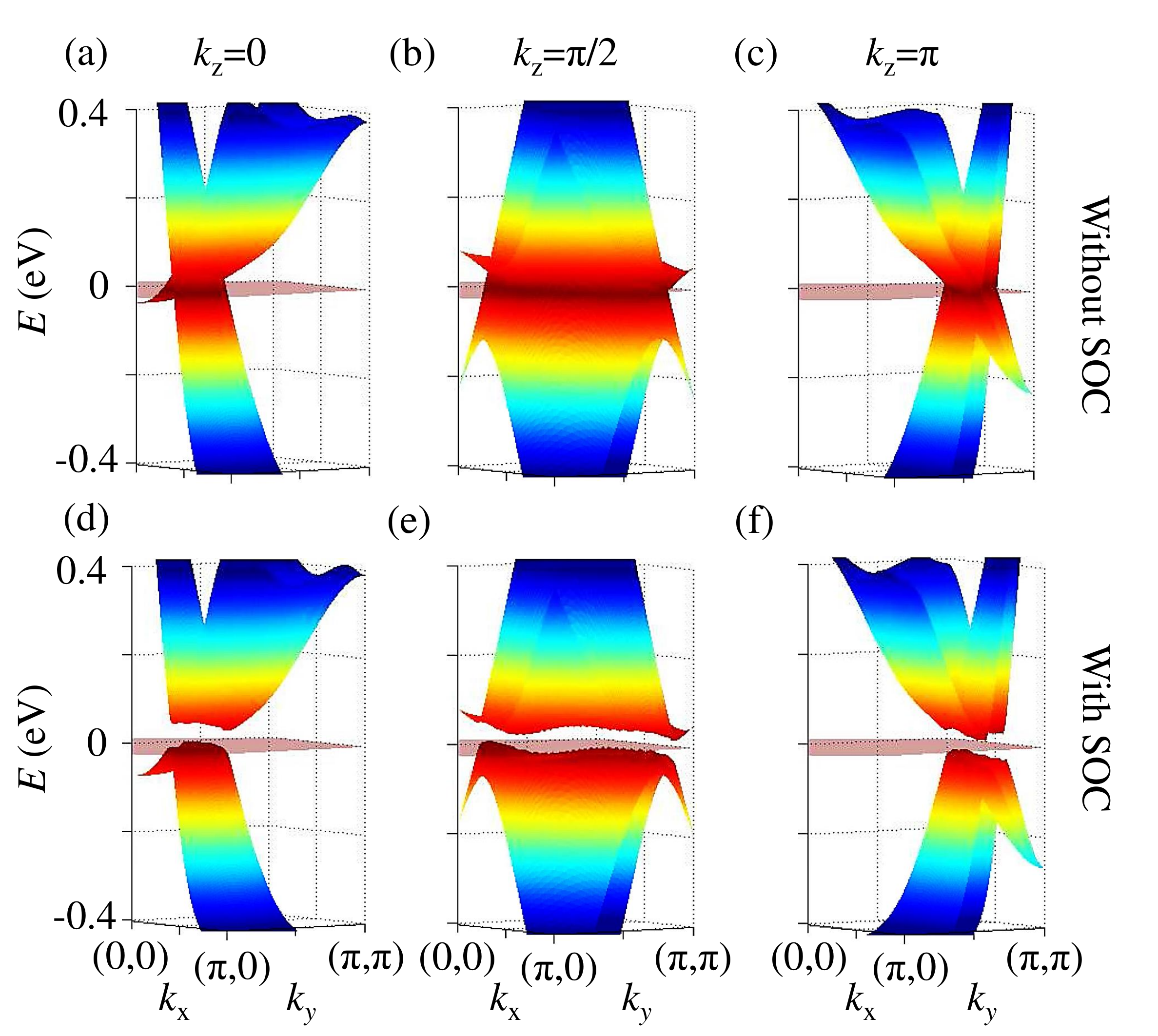}}
\vspace{10pt}\\
\parbox{\textwidth}{
Fig. S1 DFT band structure of FM LaBr without (upper panel) and with SOC (lower panel) in the $k_x-k_y$ plane for three representative $k_z$ cuts. (a-c) Without SOC, two bands coming from the majority and minority spins of $d_{x^2-y^2}$- and $d_{z^2}$-orbitals cross each other at a momentum contour on the $k_x-k_y$ plane near the Fermi level, which we call as nodal Weyl ring. The band crossing occurs at all values of $k_z$. As SOC is turned on, the Weyl ring is gapped out at all momenta, giving rise to a QAH insulator. The results are similar at all other $k_z$ cuts. The result for SOC is shown here for the 0.24\% strain case in which a fully insulating gap opens. At lower strain, although the degeneracy is lifted at momentum, tiny semimetallic pockets appears as discussed in the main text. The horizontal plane in each figure is guide to the eye to the Fermi level plane. The color gradient has no physical meaning.}
\end{figure}

\begin{figure}[h!]
\centering
\fbox{\includegraphics[scale=0.33]{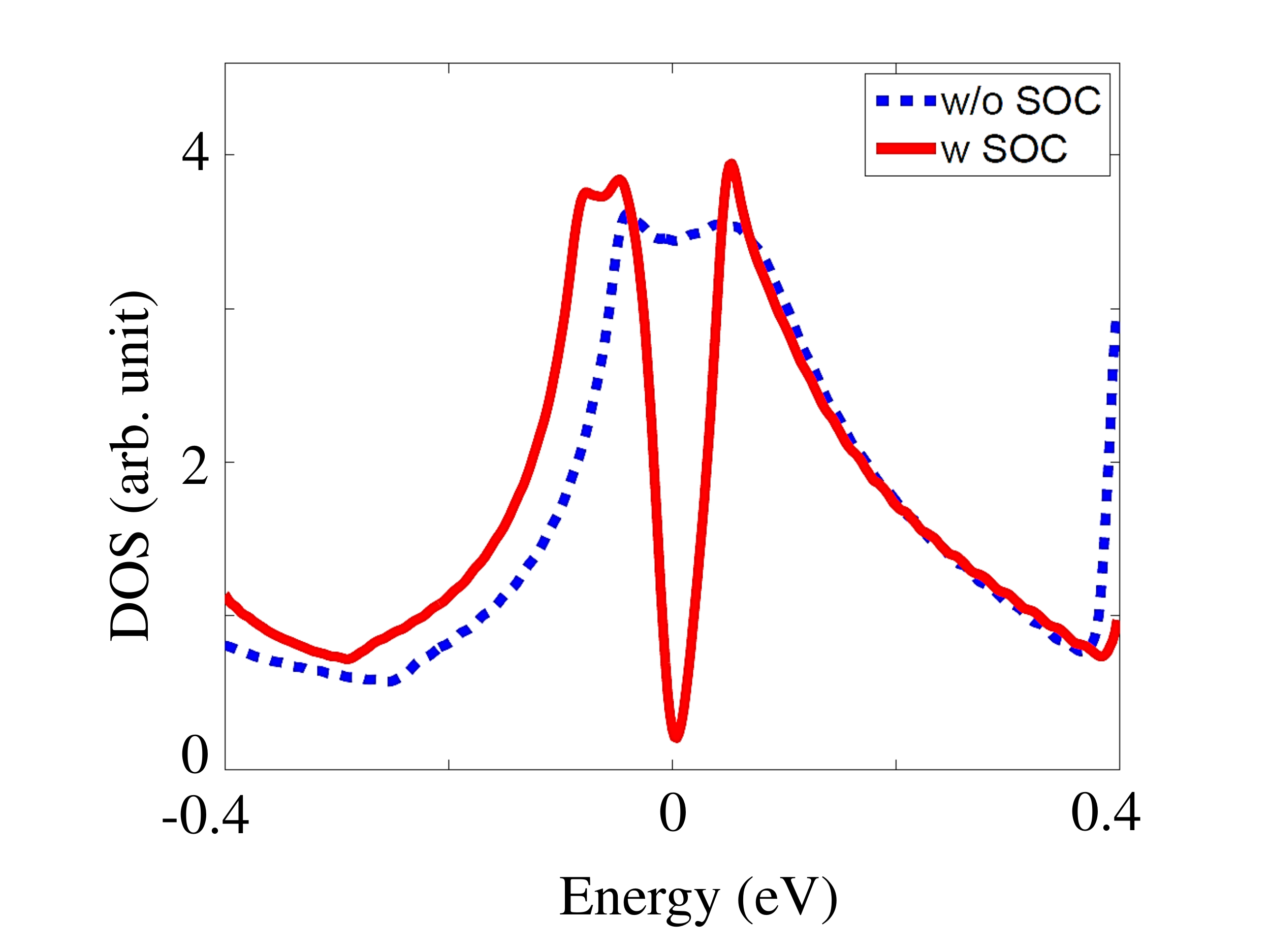}}
\vspace{10pt}\\
\parbox{\textwidth}{Fig. S2. (a) Density of states (DOS) for the same two cases as in Fig.~S1, for FM LaBr without and with SOC. With SOC, an insulating gap is opened at the Fermi level. Small value of DOS is an artifact due to the finite number of k-points and large broadening. With smaller broadening, a fully insulating gap is visible, but the overall DOS shows noisy behavior. (b) Surface state (red lines) along representative high-symmetry directions where SOC opens a gap in Fig. S1. For all cases, we have used eight La-bilayers. For $\bar{\Gamma}\rightarrow {\bar {\rm L}}$ direction, we have made the supercell along $b$-direction, with $k_x$ and $k_z$ remaining the good quantum number. Similar consideration is taken for all other directions.}
\end{figure}

\begin{figure}[h!]
\centering
\fbox{\includegraphics[scale=0.44]{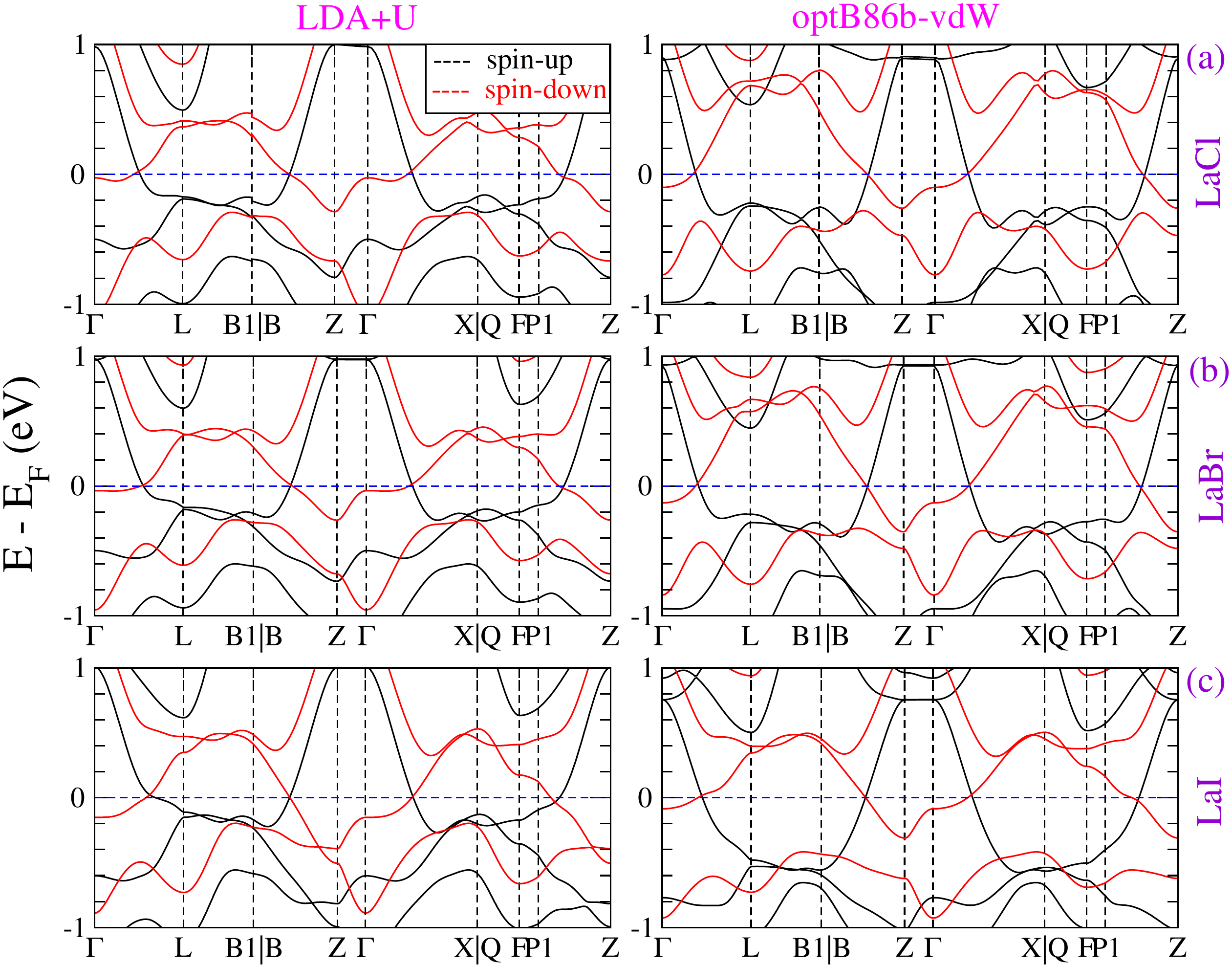}}
\fbox{\includegraphics[scale=0.4]{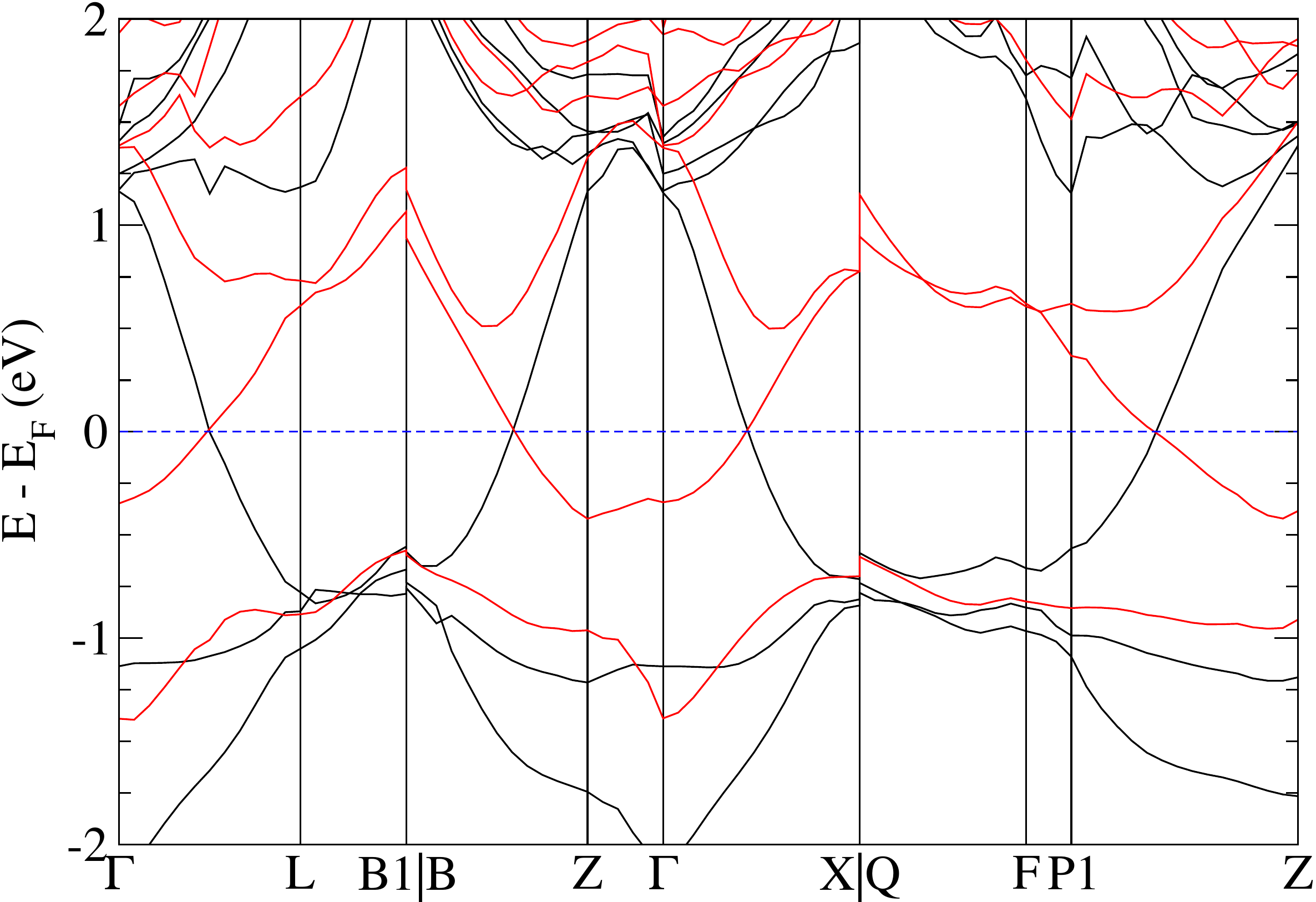}}
\vspace{12pt}\\
\parbox{\textwidth}{
 Fig. S3. (Upper box): Comparison of band structure calculated using different methods for all three materials in the same FM state. (Lower-panel) The HSE06 bandstructure of LaI without including SOC. In order to assess the robustness of  the band inversion in our computed electronic structure due to the underestimation of band gap within the semi-local functionals LDA or GGA, we calculated the bandstrructure of LaI, a representative material of the LaX class, using the hybrid functional HSE06. For all calculations, the value of $U$ is kept fixed at $U$=8~eV. The band structures are reproducible by four different functionals studied here.}
\label{method}
\end{figure}

\begin{figure}[h!]
\centering
\fbox{\includegraphics[scale=0.5]{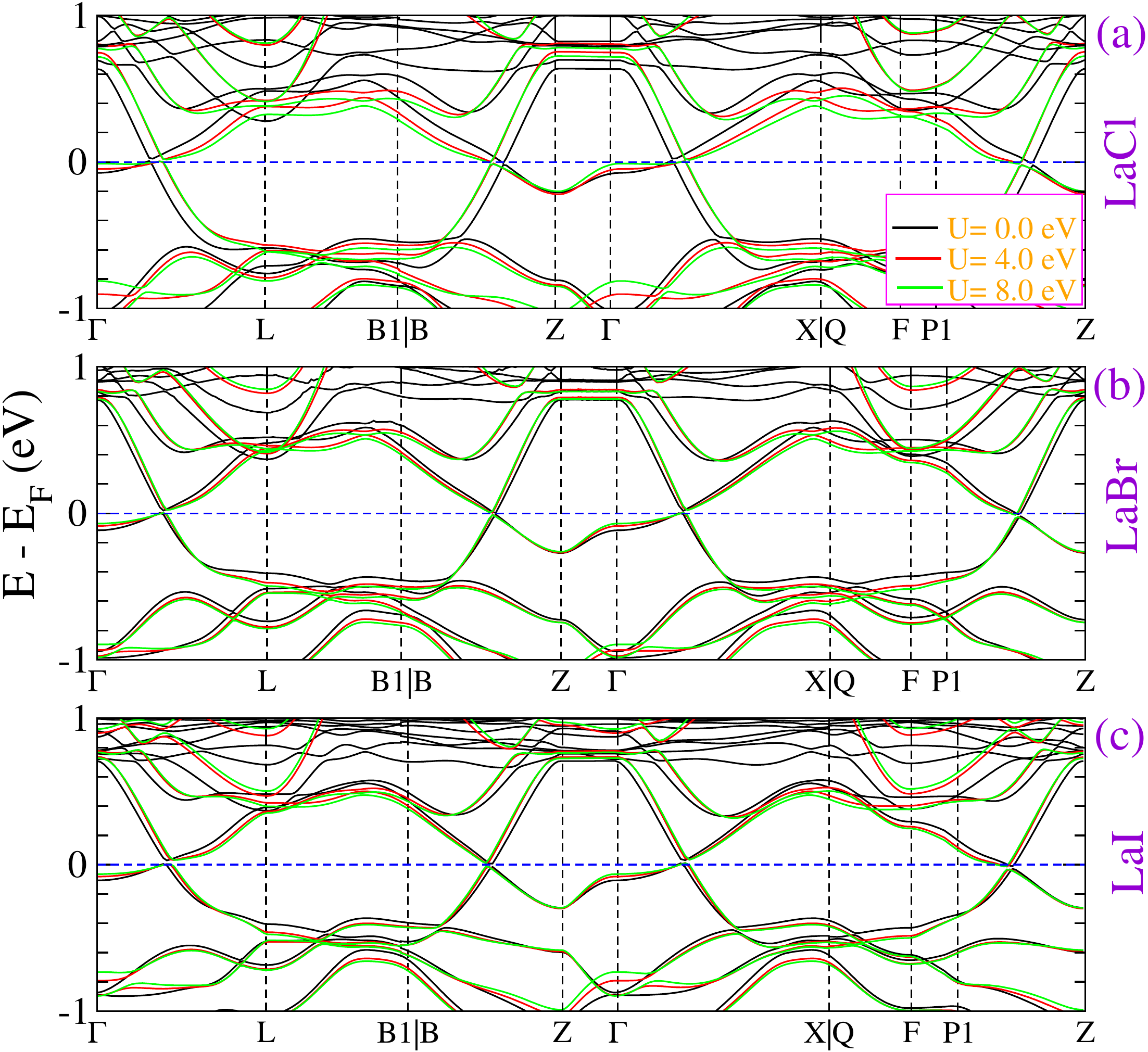}}
\vspace{12pt}\\
\parbox{\textwidth}{
 Fig. S4. GGA+U band structure using different values of $U$.}
\label{method}
\end{figure}

\begin{figure}[h!]
\centering
\fbox{\includegraphics[scale=0.5]{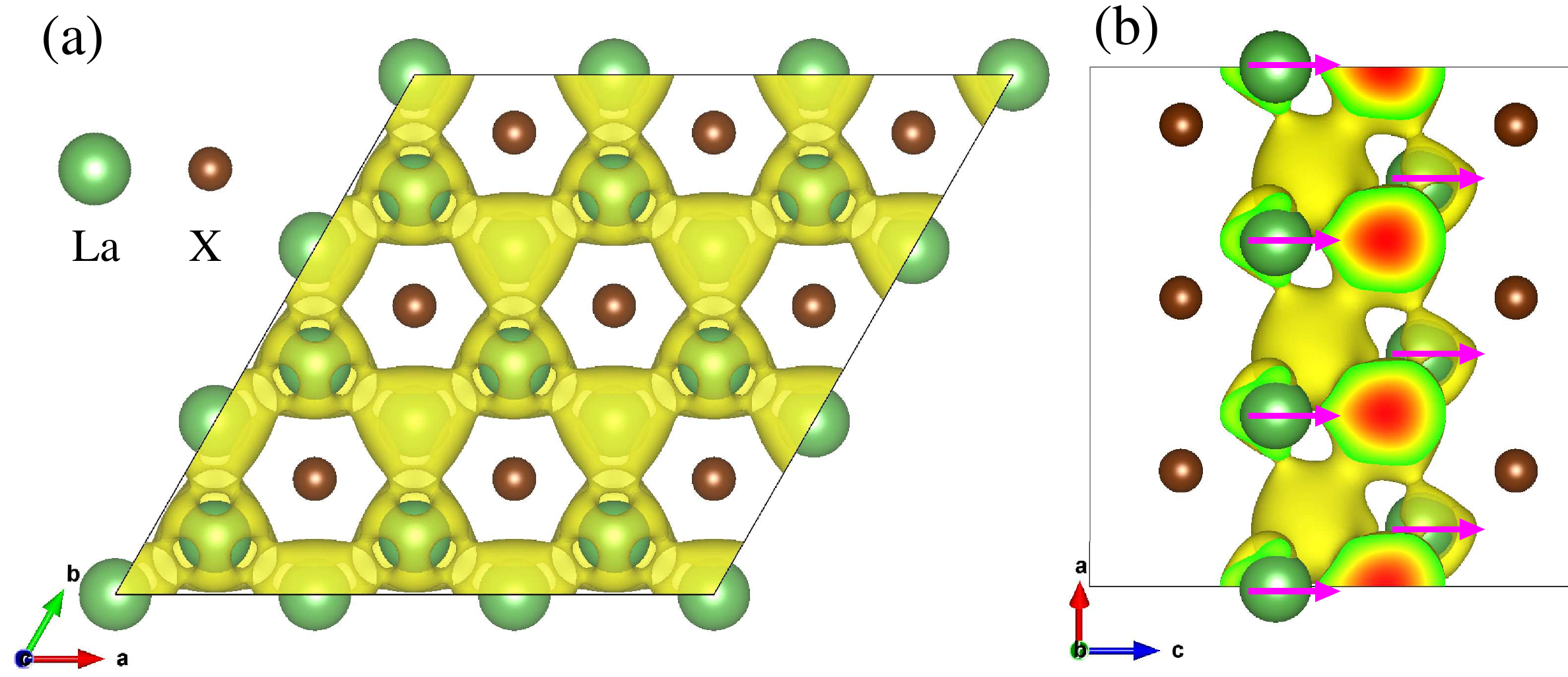}}
\vspace{12pt}\\
\parbox{\textwidth}{
Fig. S5. Spatial distribution of spin density, defined as $n_{\uparrow}-n_{\downarrow}$, where $n_{\uparrow}$ and $n_{\downarrow}$ are the total electron density of spin-up and spin-down, respectively in the FM state without SOC. Yellow color shows the isosurface at 0.0043~\AA$^3$/electrons. (a)-(b) Top and side views of the same spin-density plots. Arrows in (b) dictate the spin orientation. For visualization purpose, we plot the spin-density of a monolayer using 3$\times$3 supercell.}
\label{phonon}
\end{figure}

\begin{figure}[h!]
\centering
\fbox{\includegraphics[scale=0.5]{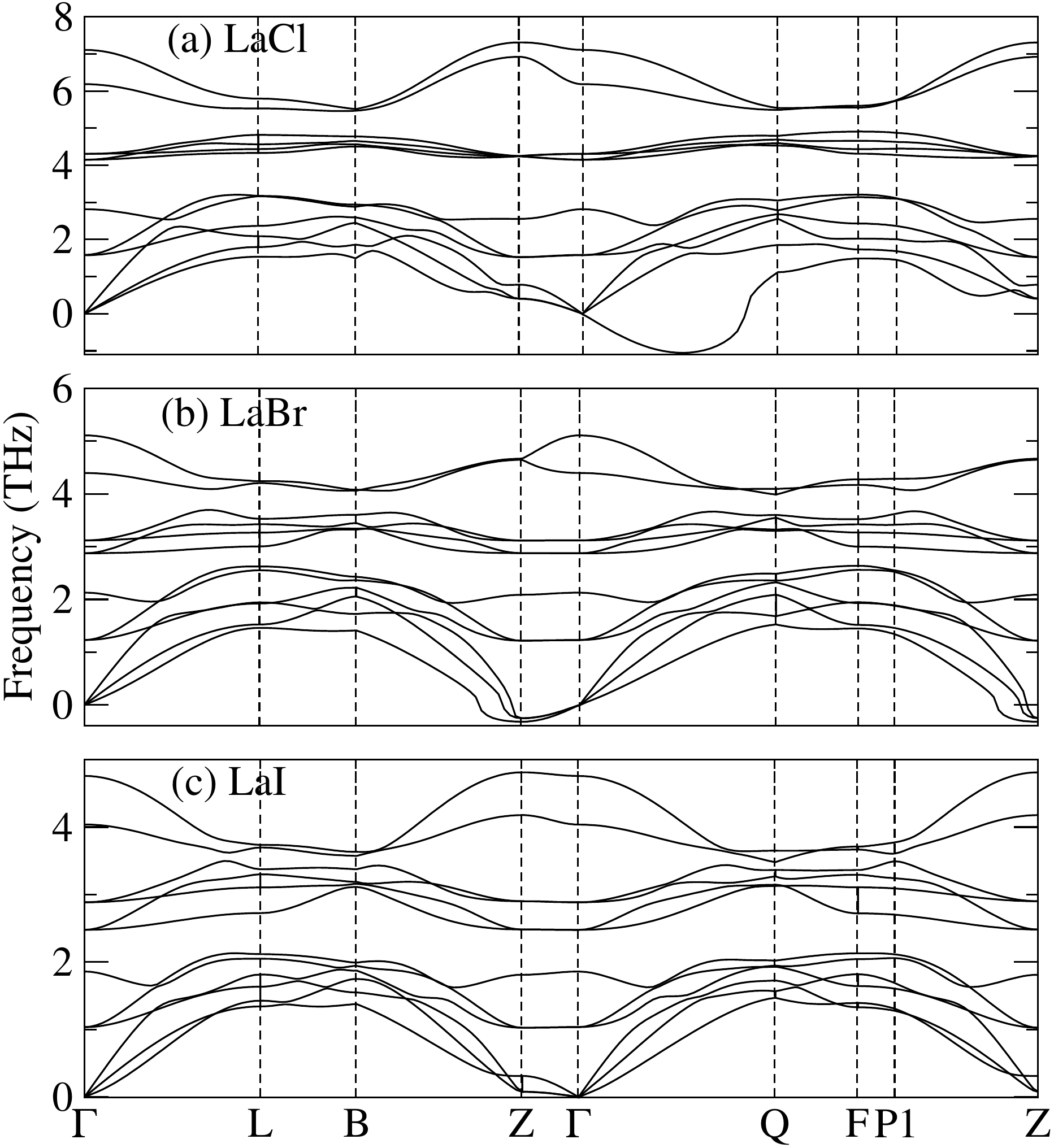}}
\vspace{12pt}\\
\parbox{\textwidth}{
 Fig. S6. Phonon band structure of  La$X$ materials. All systems exhibit positive phonon frequency at the $\Gamma$-point, implying that they have stable lattice structure. Small negative phonon frequencies at non-high-symmetric momenta for LaBr and LaCl can be elevated by small strain.}
\label{phonon}
\end{figure}

 
\noindent
\clearpage
\noindent
{\large{\bf Supporting Tables}}\\

\begin{table}[h]
\centering
\begin{tabular}{c |c |c| c}
Materials & $H_{\rm f}^{\rm DFT}$  in eV (eV/$X$)  & $H_{\rm f}^{\rm DFT}$ in  kJ mol$^{-1}$ & $H_{\rm f}^{\rm  exp}$  in kJ mol$^{-1}$ \\
\hline \hline 
LaCl     & -2.75  (-2.75) & -265.33 & -\\
LaCl$_3$ & -8.80  (-2.93) & -849.07 & -1071.6~\cite{Hf-expt} \\
LaBr     & -2.42  (-2.42) & -233.49 & - \\
LaBr$_3$ & -7.82  (-2.61) & -754.52 & -904.4~\cite{Hf-expt}\\
LaI      & -1.78  (-1.78) & -171.74 &-\\
LaI$_3$  & -5.76  (-1.92) & -555.76 & -673.9~\cite{Hf-expt}\\
\hline
\end{tabular}
\vspace{12pt}\\
\parbox{\textwidth}{
 Supporting Table SI. Theoretically estimated values of the formation enthalpies for both La$X$ and LaX$_3$ compounds. The value in bracket (2nd column) represents the formation enthalpy per $X$ atom. The hexgonal crystal structure with space group P6$_3$/m is used for La$X_3$. Note that our calculated formation enthalpy formally represents the value at zero temperature and does not account for any prohibited kinetic barriers involved in structural changes. We calculated the formation enthalpy according to the reaction $x$La(s) + y$X$(s) $\rightarrow$~La$_x$$X_y$, s corresponds to solid phase. The specific formula is given in the main text.}
\label{tab-cohesive}
\end{table}
\begin{table}[h]
\centering
\begin{tabular}{|p{2cm}| p{7cm} | c c c c|}
\hline \hline 
Material & Crystal struture & Atoms & x & y & z \\
\hline
\multicolumn{6}{|c|}{GGA + U method}\\
\hline
      & $a$=$b$=$c$ = 10.208~\AA; & La1 & 0.885 & 0.885 & 0.885 \\  
LaCl  & $\alpha$ = $\beta$ = $\gamma$ = 23.706$^{\circ}$ & La2 & 0.115 & 0.115 & 0.115 \\ 
      & & Cl1 & 0.613 & 0.613 & 0.613 \\
      & & Cl2 & 0.387 & 0.387 & 0.387 \\
\hline
      & $a$=$b$=$c$ = 10.993~\AA; & La1 & 0.880 & 0.880 & 0.880 \\  
LaBr  & $\alpha$ = $\beta$ = $\gamma$ = 21.709$^{\circ}$ & La2 & 0.120 & 0.120 & 0.120 \\ 
      & & Br1 & 0.608 & 0.608 & 0.608 \\
      & & Br2 & 0.392 & 0.392 & 0.392 \\
\hline     
      & $a$=$b$=$c$ = 11.922~\AA; & La1 & 0.877 & 0.877 & 0.877 \\  
LaI   & $\alpha$ = $\beta$ = $\gamma$ = 20.783$^{\circ}$ & La2 & 0.123 & 0.123 & 0.123 \\ 
      & & Br1 & 0.608 & 0.608 & 0.608 \\
      & & Br2 & 0.392 & 0.392 & 0.392 \\ 
\hline
\multicolumn{6}{|c|}{LDA + U method}\\
\hline
      & $a$=$b$=$c$ = 9.590~\AA; & La1 & 0.888 & 0.888 & 0.888 \\  
LaCl  & $\alpha$ = $\beta$ = $\gamma$ = 24.359$^{\circ}$ & La2 & 0.112 & 0.112 & 0.112 \\ 
      & & Cl1 & 0.619 & 0.619 & 0.619 \\
      & & Cl2 & 0.381 & 0.381 & 0.381 \\
\hline
      & $a$=$b$=$c$ = 10.213~\AA; & La1 & 0.885 & 0.885 & 0.885 \\  
LaBr  & $\alpha$ = $\beta$ = $\gamma$ = 22.983$^{\circ}$ & La2 & 0.115 & 0.115 & 0.115 \\ 
      & & Br1 & 0.618 & 0.618 & 0.618 \\
      & & Br2 & 0.382 & 0.382 & 0.382 \\
\hline     
      & $a$=$b$=$c$ = 11.506~\AA; & La1 & 0.877 & 0.877 & 0.877 \\  
LaI   & $\alpha$ = $\beta$ = $\gamma$ = 20.660$^{\circ}$ & La2 & 0.123 & 0.123 & 0.123 \\ 
      & & Br1 & 0.608 & 0.608 & 0.608\\
      & & Br2 & 0.392 & 0.392 & 0.392 \\ 
\hline
\multicolumn{6}{|c|}{vdW-DFT method}\\
\hline
      & $a$=$b$=$c$ = 9.527~\AA; & La1 & 0.885 & 0.885 & 0.885 \\  
LaCl  & $\alpha$ = $\beta$ = $\gamma$ = 24.311$^{\circ}$ & La2 & 0.115 & 0.115 & 0.115 \\ 
      & & Cl1 & 0.615 & 0.615 & 0.615 \\
      & & Cl2 & 0.385 & 0.385 & 0.385 \\
\hline
      & $a$=$b$=$c$ = 10.373~\AA; & La1 & 0.882 & 0.882 & 0.882 \\  
LaBr  & $\alpha$ = $\beta$ = $\gamma$ = 22.467$^{\circ}$ & La2 & 0.118 & 0.118 & 0.118 \\ 
      & & Br1 & 0.614 & 0.614 & 0.614 \\
      & & Br2 & 0.386 & 0.386 & 0.386 \\
\hline     
      & $a$=$b$=$c$ = 11.952~\AA; & La1 & 0.877 & 0.877 & 0.877 \\  
LaI   & $\alpha$ = $\beta$ = $\gamma$ = 20.696$^{\circ}$ & La2 & 0.123 & 0.123 & 0.123 \\ 
      & & Br1 & 0.608 & 0.608 & 0.608 \\
      & & Br2 & 0.392 & 0.392 & 0.392 \\ 
\hline
\end{tabular}
\vspace{12pt}\\
\parbox{\textwidth}{
Supporting Table SII. Relaxed crystal structure and atomic coordinates of different La$X$ ($X$=Cl, Br and I) compounds for different methods.}
\label{tab-structure-GGAU}
\end{table}

\end{widetext}

\end{document}